\documentstyle[12pt]{article}
\begin{document}

\title{\LARGE\bf Trace anomaly in the field-antifield formalism}

\author{J. Barcelos-Neto\thanks{\noindent e-mails: ift03001
@ ufrj and barcelos @ vms1.nce.ufrj.br}, N.R.F.
Braga\thanks{\noindent e-mail:braga@vms1.nce.ufrj.br} and S.M. de Souza\\
Instituto de F\'{\i}sica\\
Universidade Federal do Rio de Janeiro\\
RJ 21945-970 - Caixa Postal 68528 - Brasil\\
\date{}}

\maketitle
\abstract
The field-antifield quantization method is used to calculate the
trace anomaly for a massless scalar field in a curved background,
by means of the zeta function regularization procedure.

\vfill
\vspace{1cm}
\newpage

{\bf 1.$\,\,$ }The quantization method of Batalin and
Vilkovisky~(BV)~\cite{BV1,BV2,He,GPS}
is considered to be the most powerful
quantization method for theories with first-class
constraints~\cite{Dirac}. It is a Lagrangian procedure and,
consequently, it is naturally covariant.  The quantum action itself
is the generator of the BRST transformations, in an extended space
that involves the fields and the associated antifields (with opposite
statistics).
\medskip
One of the main advantages of the BV method, besides the natural
covariance, is that it can be used for gauge theories whose
generators form an open algebra \cite{BV2}. In this case, the
 {\it
ghost-for-ghost} structure, characteristic of reducible
systems~\cite{He}, is build up in a very
simple way.

\medskip
In addition, recent
results \cite{GPS} show that the application of the BV formalism to anomalous
gauge theories lead to a very systematic and elegant mechanism of
calculating the anomalies associated to the non trivial
behavior of the path integral measure
and also of generating the Wess Zumino terms that compensate this
contributions, leading to a BRST invariant theory.  In other words,
 the BV method realizes, in a very natural procedure, the mechanism
proposed by Faddeev and Shatashvili of
restoring gauge invariance, in the BRST language.
\medskip

The purpose of the present work refers just to this last case. We use the
BV method to obtain the trace anomaly for a massless scalar field
moving in a curved background~\cite{Bi}. It is important to emphasize
that when the BV method is used in the computation of anomalies it
must be accompanied with some regularization procedure~\cite{TPN}.
Usually one uses the Pauli-Villars regularization method when dealing
with  BV quantization. We will use here the zeta function method, that
can be naturally formulated in curved space.

\bigskip
{\bf 2.$\,\,$ } The BV quantization procedure is defined in an
enlarged space of fields and antifields, collectively denoted by
$\Phi^a$ and $\Phi^{*a}$ respectively.  The quantum action has the
general $\hbar$ expansion :

\begin{equation}
\label{action}
W(\Phi^a ,\Phi^{*a} ) =
S(\Phi^a ,\Phi^{*a} ) +
\sum_{p=1}^\infty \hbar^p M_p (\Phi^a ,\Phi^{*a} ) \, .
\end{equation}

\bigskip
The zero order term of the action $W$ : $  S (\Phi^a,\Phi^{\ast a})$
is usually called gauge fixed action  and is subject to the boundary
condition:

\begin{equation}
\label{boundary}
 S(\Phi^a,\Phi^{\ast a}= 0)= {\cal S} (\phi^{i} )\, ,
\end{equation}
\bigskip
\noindent where ${\cal S}$ is the action of the classical theory.
\medskip
The first order term $M_1$ corresponds to one loop quantum corrections.
It has recently  been shown that it is possible to  enlarge
the space of fields and antifields in such a way that $M_1\,$
plays the role of Wess Zumino term, restoring the BRST invariance
of the theory\cite{GPS,BM}.
\medskip
The condition of BRST invariance of the quantum theory is
translated in the BV formalism to the condition that the so called
master equation must be satisfied:

\begin{equation}
\label{Master}
{1\over 2}(W,W) = i\hbar\Delta W
\end{equation}

\bigskip
\noindent  where the antibracket of two generic functions $X$ and $Y$
is defined as:

\begin{equation}
(X,Y) = {\partial_rX\over
\partial\Phi^a} {\partial_lY\over\partial\Phi^{\ast a}}
- {\partial_rX\over \partial\Phi^{\ast a}}
  {\partial_lY\over \partial\Phi^a}
\end{equation}
\bigskip
\noindent and the operator $\Delta$ as:
\begin{equation}
\Delta \equiv
{\partial_r\over\partial\Phi^a}{\partial_l\over\partial\Phi^\ast_a}\;
\end{equation}
\bigskip
Anomalies correspond just to a violation in the master equation that,
after a suitable regularization procedure is applied, can be written
in the form \cite{TPN}:

\begin{equation}
\label{Anomaly}
{1\over 2}(W,W)- i\hbar\Delta W = \int d^4x \, C^\gamma A_\gamma \, .
\end{equation}

\bigskip
\noindent The symmetries associated to the ghosts $C^\gamma$
are said to be broken at the quantum level.
More details about the field-antifield quantization method can be
found in recent reviews available in the literature~\cite{He,GPS}.

\bigskip
{\bf 3.$\,\,$ } Let us  consider a classical action representing a
 scalar field in a curved spacetime

\begin{equation}
{\cal S} =\frac{1}{2}\int d^4x\,\sqrt{-g}\,\,\Bigl(g_{\mu\nu}\,
\partial^\mu\phi\,\partial^\nu\phi
-\frac{1}{6}\,R\,\phi^2\Bigr)\,\, ,
\label{2.1}
\end{equation}

\bigskip\noindent
where $g_{\mu\nu}$ is the metric tensor and $R$ is the
curvature scalar . The Lagrangian above is invariant under the Weyl
(conformal) transformations

\begin{eqnarray}
\label{Weyl}
\delta g_{\mu\nu}(x)&=&2\,\alpha(x)\,g_{\mu\nu}\,\, ,
\nonumber\\
\delta\phi(x)&=&-\, \alpha(x)\,\phi(x)\,\, ,
\label{2.2}
\end{eqnarray}
\bigskip
\noindent that leads to a vanishing trace for the energy-momentum tensor.
Here, $\alpha(x)$ is some (infinitesimal)
continuous and real function. It is well-known that this symmetry is
not preserved in the quantum scenario~\cite{Bi,Fu}. This corresponds
to the so-called Weyl anomaly, where the trace of the energy momentum
tensor is not zero anymore.

\medskip
Our purpose is to calculate the anomaly, and not to find a BRST invariant
representation for the theory, therefore
we will consider that all the $ M_p$ terms of the quantum action $W$
(eq. (\ref{action})) can be taken as zero.  In this case, equation
(\ref{Anomaly}) becomes

\begin{equation}
\label{Anomaly2}
- i\hbar\Delta S = \int d^4x\,\sqrt{-g}\,\, C\,\, A \,\,\,\, ,
\end{equation}
\bigskip
\noindent where the $\sqrt{-g}$ factor was included in the volume integral
since we are dealing with non flat space.
\medskip
First we build up the  gauge fixed action $S$ following the standard
procedure ~\cite{BV1,He} getting:

\begin{equation}
S=\int d^4x\,\sqrt{-g}\,\Bigl(\frac{1}{2}\,
g_{\mu\nu}\,\partial^\mu\phi\,\partial^\nu\phi
-\frac{1}{12}\,R\,\phi^2
-\phi^\ast\,C\,\phi\Bigr)\,\, ,
\label{2.3}
\end{equation}
\bigskip
\noindent where $C$ is the ghost field associated to the symmetry
(\ref{Weyl}) and, as usual,
antifields are denoted by a star superscript.
It is important to remark that the metric $g_{\mu\nu}$ is taken as
an external background field ( not as a quantum field) thus there
is no antifield associated to it in $S$.
\medskip
Now we must calculate:

\begin{equation}
\Delta\,S=\frac{\delta^R}{\delta\phi(x)}\,
\frac{\delta^L}{\delta\phi^\star(x)}\,S\,\, .
\label{2.4}
\end{equation}
\bigskip
\noindent As it is discussed in \cite{TPN}, this expression is not well
defined as it would be proportional to a $\delta(0)$. One must
regularize the action $S$ before acting the $\Delta$ operator on it.
We will consider the zeta function regularization procedure~\cite{Ze}.

\medskip
Let us consider a complete and orthonormal set of eigenfunctions
$\phi_n$ of the operator that appears in the action
(\ref{2.1}), that is to say

\begin{equation}
\Bigl(\Box^2+\frac{1}{6}\,\Bigr)\,\phi_n=\lambda_n\,\phi_n\,\, ,
\label{2.5}
\end{equation}

\bigskip
\noindent with

\begin{eqnarray}
\int d^4x\,\sqrt{-g}\,\phi_n(x)\,\phi_m(x)&=&\delta_{nm}\,\, ,
\label{2.6}\\
\sum_m\phi_m(x)\,\phi_m(y)
&=&\frac{\delta(x-y)}{\sqrt{-g}}\,\, .
\label{2.7}
\end{eqnarray}
\bigskip
\noindent We introduce this regularization in the last term of $S$,
by writing it as:

\begin{eqnarray}
& &\int d^4x\,\sqrt{-g}\,
\phi^\ast (x)\,C (x)\,\phi (x) =
 \int d^4y \, d^4x\, \sqrt{-g}\, \phi^\ast (x)\,C (x)\, \delta^4(x-y)
\,\phi (y)\nonumber\\
&=& \int d^4y \, d^4x\, \sqrt{-g} \, \phi^\ast (x)\,C (x)\,
\sum_m\phi_m(x)\,\phi_m(y) \sqrt{-g}\,
\phi (y)\,\, .
\label{2.8}
\end{eqnarray}
\bigskip
\noindent We get thus:

\begin{equation}
\Delta\,S= - \int d^4x \sqrt{-g}\,C \,\sum_m\phi_m(x)
\,\phi_m(x)\,\, .
\label{2.9}
   \end{equation}

\vspace{1cm}

The  previously mentioned illness in
$\Delta S$ is now contained in the sum \break
$\sum_m\phi_m(x) \,\phi_m(x)$
above. We regularize it by using the zeta function technique in
curved spacetime~\cite{Bi,Ze}. The generalized zeta function
associated with the operator of eq.~({2.5}) is

\begin{equation}
\zeta(s,x)=\sum_m\frac{\phi_m(x)\,\phi_m(x)}{\lambda^s_m}
\label{2.11}
\end{equation}

\bigskip
\noindent
We thus have that the sum in equation (\ref{2.9}) can be written
in terms of the zeta function as

\begin{equation}
\sum_m\phi_m(x)\,\phi_m(x)=\lim_{s\rightarrow0}\zeta(s,x)\,\, .
\label{2.12}
\end{equation}

\bigskip\noindent
To evaluate the zeta function $\zeta(s,x)$ at $s=0$ we use the heat
equation associated with the operator of eq.~(\ref{2.6}), that is

\begin{equation}
\frac{d}{d\tau}\,K(x,y,\tau)
+\Bigl(\Box^2+\frac{1}{6}\Bigr)\,
K(x,y,\tau)=0
\label{2.12a}\,\, ,
\end{equation}

\bigskip\noindent
with the condition

\begin{equation}
K(x,y,0)=\frac{\delta^{(4)}(x-y)}{\sqrt{-g}}\,\, ,
\label{2.13}
\end{equation}

\bigskip\noindent
where $\tau$ is a proper time evolution parameter. It is easy to see
that the function $K(x,y,\tau)$ has the form

\begin{equation}
K(x,y,\tau)=\sum_me^{-\lambda_m\tau}\,\phi_m(x)\,\phi_m(y)\,\, .
\label{2.14}
\end{equation}

\bigskip\noindent
It is also straightforward to express the generalized zeta function
in terms of the heat kernel,

\begin{eqnarray}
\zeta(x,s)&=&\frac{1}{\Gamma(s)}\int_0^\infty d\tau\,
\tau^{s-1}\,K(x,x,\tau)
\nonumber\\
&=&\frac{1}{16\pi^2\Gamma(s)}\int_0^\infty d\tau\,
\tau^{s-3}\,\Omega(x,x,\tau)\,\, ,
\label{2.15}
\end{eqnarray}

\bigskip\noindent
where

\begin{equation}
K(x,x,\tau)=\frac{1}{(4\pi\tau)^2}\,\Omega(x,x,\tau)\,\, .
\label{2.16}
\end{equation}

\bigskip
It is assumed that $\Omega(x,x,\tau)$ and their derivatives are
dumped for large $\tau$. Thus, from (\ref{2.15}) we obtain

\begin{equation}
\zeta(x,0)=\frac{1}{32\pi^2}\,
\frac{\partial^2\Omega(x,x,\tau)}{\partial\tau^2}\,
\vert_{\tau=0}
\label{2.17}
\end{equation}

\bigskip\noindent
For small $x$, one also assumes the following expansion for
$\Omega(x,x,\tau)$

\begin{equation}
\Omega(x,x,\tau)\sim a_0(x)+a_1(x)\,\tau
+a_2(x)\,\tau^2+0(\tau^3)
\label{2.18}
\end{equation}

\bigskip\noindent
where the coefficients $a_m(x)$ are dimensionally independent and the
first few of them have been calculated for some kind of Hermitian
operators using the method of coincidence limit~\cite{Witt}. The
combination of (\ref{2.12}), (\ref{2.17}) and (\ref{2.18}) gives

\begin{equation}
\Delta S=-\int d^4x\,\sqrt{-g}\,C\,
\frac{a_2}{16\pi^2}
\label{2.19}
\end{equation}

\bigskip
\noindent equation (\ref{Anomaly2}) then tells us that the anomaly is

\begin{equation}
\label{Anomaly3}
A =  i \hbar {a_2\over 16\pi^2} \, ,
\end{equation}

\bigskip\noindent
The coefficient function $a_2(x)$ for the operator
(\ref{2.5}) is calculated in a straightforward way~\cite{Witt}. The
result is

\begin{equation}
a_2=\frac{1}{180}\,
\Bigl(R_{\mu\nu\xi\rho}\,R^{\mu\nu\xi\rho}
-R_{\mu\nu}\,R^{\mu\nu}-\Box R\Bigr)
\label{2.20}
\end{equation}

\bigskip
\noindent This result is in agreement with previous calculations
\cite{Bi,Fu}.
\bigskip
Concluding, we have shown, through a typical example,
how one can  use the field anti-field formalism in order
to calculate the trace anomaly.  Our calculations provide
an example of the use of an alternative regularization
technique ( the zeta  function )
in the  BV formalism, in contrast with the usual
Pauli Villars regularization found in the literature \cite{GPS}.

\vspace{1cm}

\vspace{1cm}
\noindent {\bf Acknowledgment:} This work is supported in part by
Conselho Nacional de Desenvolvimento Cient\'{\i}fico e Tecnol\'ogico
- CNPq (Brazilian Research Agency).

\end{document}